\documentclass
[superscriptaddress,secnumarabic,amssymb,amsmath,nobibnotes,aps,prd,showkeys,showpacs,nofootinbib]{revtex4}%
\usepackage{graphicx}
\usepackage{epsf}
\usepackage{bm}
\usepackage{amsmath}
\usepackage{amsfonts}
\usepackage{amssymb}
\usepackage{epstopdf}%
\providecommand{\U}[1]{\protect\rule{.1in}{.1in}}
\newcommand{\be}{\begin{equation}}
\newcommand{\ee}{\end{equation}}

\newcommand{\mincir}{\raise
-3.truept\hbox{\rlap{\hbox{$\sim$}}\raise4.truept\hbox{$<$}\ }}
\newcommand{\magcir}{\raise
-3.truept\hbox{\rlap{\hbox{$\sim$}}\raise4.truept\hbox{$>$}\ }}

\begin{document}
\title{Dynamical symmetries and observational constraints in scalar field cosmology}
\author{Andronikos Paliathanasis}
\email{paliathanasis@na.infn.it}
\affiliation{Dipartimento di Fisica, Universita' di Napoli, ``Federico II'', Complesso
Universitario di Monte S. Angelo, Edificio G, Via Cinthia, I-80126, Napoli, Italy}
\affiliation{INFN, Sezione di Napoli, Complesso Universitario di Monte S. Angelo, Via
Cintia Edificio 6, I-80126 Napoli, Italy}
\author{Michael Tsamparlis}
\email{mtsampa@phys.uoa.gr}
\affiliation{Faculty of Physics, Department of Astrophysics - Astronomy - Mechanics
University of Athens, Panepistemiopolis, Athens 157 83, Greece}
\author{Spyros Basilakos}
\email{svasil@academyofathens.gr}
\affiliation{Academy of Athens, Research Center for Astronomy and Applied Mathematics,
Soranou Efesiou 4, 11527, Athens, Greece}

\begin{abstract}
We propose to use dynamical symmetries of the field equations, in order to
classify the dark energy models in the context of scalar field (quintessence
or phantom) FLRW cosmologies. Practically, symmetries provide a useful
mathematical tool in physical problems since they can be used to simplify a
given system of differential equations as well as to determine the
integrability of the physical system. The requirement that the field equations
admit dynamical symmetries results in two potentials one of which is the well
known \textit{Unified Dark Matter} (UDM) potential and another new potential.
For each hyperbolic potential we obtain the corresponding analytic solution of
the field equations. The proposed analysis suggests that the requirement of
the contact symmetry appears to be very competitive to other independent tests
used to probe the functional form of a given potential and thus the associated
nature of dark energy. Finally, in order to test the viability of the above
scalar field models we perform a joint likelihood analysis using some of the
latest cosmological data.
%SNIa, BAO and H(z) data.

\end{abstract}

\pacs{98.80.-k, 95.35.+d, 95.36.+x}
\keywords{Cosmology; dark energy; scalar field; dynamical symmetries}\maketitle

\section{Introduction}

The detailed analysis of the current cosmological data indicate that the
universe is spatially flat and has incorporated two acceleration phases. An
early acceleration phase (inflation), which occurred prior the radiation
dominated era and a recently initiated accelerated expansion (see
\cite{Teg04,Spergel07,essence,Kowal08,Hic09,komatsu08,LJC09,BasPli10,komatsu11,Ade13}
and references therein). The source for the late time cosmic acceleration has
been attributed to an unidentified type of \textquotedblright
matter\textquotedblright\ with negative equation of state, usually called dark
energy (DE). Despite the mounting observational evidences on the existence of
the DE component in the universe, its nature has yet to be found (for a review
see \cite{Ame10} and references therein).

The easiest path for DE corresponds to the so called cosmological constant
(see \cite{Weinberg89,Peebles03,Pad03} for reviews). Indeed the spatially flat
concordance $\Lambda$CDM model, which contains cold dark matter (DM) and a
cosmological constant $\Lambda,$ fits accurately the current cosmological data
and thus it is an excellent candidate as a model which describes the observed
universe. However, more complex dynamics are necessary since the idea of a
rigid cosmological constant or vacuum energy is very difficult to reconcile
with a possible solution of the cosmological constant (tuning and coincidence
problems) plaguing theoretical cosmology \cite{Weinberg89,coincidence}. A
constant cosmological constant term throughout the entire history of the
universe presents strong conceptual difficulties from the point of view of
fundamental physics.

Attempts to overcome the above cosmological problems have been presented in
the literature (see \cite{Peebles03,Pad03,Egan08} and references therein), by
replacing the constant vacuum energy with a DE that evolves with time. Popular
proposals for the DE are, among others, the existence of new fields in nature
and the modified gravity (see
\cite{Ratra88,Oze87,Chen90,Carvalho92,Lima94,Bas09c,Caldwell98,KAM,Caldwell,Bento03,chime04,Linder2004,LSS08,SR,Xin,SVJ,Basi09}
and references therein). Particular attention over the last decades has been
paid on scalar field DE \cite{Ame10} due to its simplicity. In the scalar
field models\thinspace\ \cite{Dolgov82} and later in the quintessence context,
one can ad-hoc introduce an adjusting or tracker scalar field $\phi
~$\thinspace\cite{Caldwell98},
%(different from the usual SM Higgs field),
rolling down the potential energy $V(\phi)$, which could mimic the DE
\cite{Peebles03,Pad03,SR,Xin,SVJ}. The potential $V(\phi)$ is not known and
one must introduce it by some kind of ad hoc assumption. There have been many
such proposals as to the form of this potential e.g. power law, hyperbolic,
exponential etc \cite{Ber07,Frieman95,Steinh,Bar00,Sahni}. However one would
like to have a fundamental method according to which one would fix the form
(or forms) of the potential. One such method is the geometric requirement that
the resulting field equations admit Noether point symmetries \cite{deRitis89}.

%"selects" the dark energy models.
In fact the idea to use Noether symmetries as a cosmological tool is not new
in this kind of studies. It has been proposed that the Noether point symmetry
approach as a selection rule for the dark energy models is a geometric
criterion; that is, the geometry of the field equations can be used as a
selection criterion in order to discriminate the dark energy models.
Specifically, such a selection approach in the framework of scalar field
cosmology has been considered in
\cite{Cap96,TsamGRG,Basilakos,capPhantom,Vak,Leach,Pal2SF} and in the context
of modified theories of gravity in
\cite{Paliathanasis,HaoW,CapHam,deSouza,Kucu,CapfR,vakfR,Dong,BasFT,PalFT,PalST}%
. Dynamically speaking, Noether symmetries are considered to play a central
role in physical problems because they provide first integrals which can be
utilized in order to simplify a given system of differential equations and
thus to determine the integrability of the system. Indeed, in \cite{TsamGRG}
it has been shown that the Lie point symmetries of a dynamical system are
related to the geometry of the underlying space where the motion occurs (a
similar analysis can be found in \cite{Aminova2000,PrinceGE,JGP}).

In the current article we attempt to generalize our previous work of
Basilakos, Paliathanasis \& Tsamparlis \cite{Basilakos} (see also
\cite{Pal2SF,Paliathanasis,PalST}) in the sense that we use dynamical Noether
symmetries instead of point Noether symmetries to select the potential of the
scalar field cosmology in a spatially flat Friedmann-Robertson-Walker
spacetime (FRW). Geometrically speaking, the Noether point symmetries of the
Lagrangian are connected with the homothetic algebra of the minisuperspace
(see \cite{TsamGRG} and references therein), and the dynamical Noether
symmetries are related with the Killing tensors of the minisuperspace
\cite{Kalotas}. Obviously, the latter implies that the Noether approach
provides a useful tool in order to study the geometrical properties of the
Lagrangian in the context of the scalar field Cosmology. I this respect, we
would like to emphasize that dynamical symmetries have properties which are
well above the corresponding properties of point symmetries. Indeed the
dynamical Noether symmetries provide conserved quantities both in Newtonian
physics and in General Relativity which point symmetries cannot. For instance,
the well known Runge-Lenz vector field of the Kepler potential \cite{OConnell}%
, the Ermakov integral \cite{Lewis,Ermakov}, and the Carter constant in the
Kerr spacetime \cite{Carter} all follow from dynamical symmetries and not from
point symmetries. These integrals are not linear in the momentum; that is,
dynamical Noether symmetries provide new conservation laws in contrast to
Noether point symmetries which give integrals linear in the momentum
\cite{Stephani,Bluman}. Furthermore, the integrals they provide contain a
larger number of degrees of freedom allowing the consideration of more
scenarios in a given dynamical problem.

To our view it is important to consider the possibility of dynamical
symmetries in scalar field cosmology. As it will be shown bellow such
symmetries (at the level of contact symmetries) exist for some hyperbolic
scalar field potentials which provide us with a wide range of possibilities.
The structure of the article is as follows. In section \ref{SFC} we review
briefly the basic elements of scalar field cosmology. In section \ref{LBsym}
we give the basic definitions of generalized symmetries. In section
\ref{DynNSF} we apply the dynamical symmetry condition and classify the
potentials of the scalar field cosmology which admit contact Noether
symmetries. In section \ref{AnS} we apply the results of the section
\ref{DynNSF} and determine the analytical solution for each model. In order to
test the viability of the resulting cosmological models in section
\ref{Cconstrain} we perform a joint likelihood analysis using some of the
latest cosmological data namely, Supernovae type Ia data (SNIa), Baryonic
Acoustic Oscillations (BAO) and the $H\left(  z\right)  $ data. Finally, the
main conclusions are summarized in section \ref{Conclusion}.

\section{Field Equations}

\label{SFC}

The scalar field contribution to the curvature of space-time can be absorbed
in Einstein's field equations as follows:
\begin{equation}
R_{\mu\nu}-\frac{1}{2}g_{\mu\nu}R=k\ \tilde{T}_{\mu\nu}\;\;\;\;\;k=8\pi G
\label{EE}%
\end{equation}
where $R_{\mu\nu}$ is the Ricci tensor and $\tilde{T}_{\mu\nu}$ is the total
energy momentum tensor given by $\tilde{T}_{\mu\nu}\equiv T_{\mu\nu}+T_{\mu
\nu}(\phi)$. Here $T_{\mu\nu}(\phi)$ is the energy-momentum tensor associated
with the scalar field $\phi$, and $T_{\mu\nu}$ is the energy-momentum tensor
of matter and radiation. Modeling the expanding universe as a perfect fluid
that includes radiation, matter and DE with $4-$velocity $U_{\mu}$, we have
$\tilde{T}_{\mu\nu}=-P\,g_{\mu\nu}+(\rho+P)U_{\mu}U_{\nu}$, where $\rho
=\rho_{m}+\rho_{\phi}$ and $P=P_{m}+P_{\phi}$ are the total energy density and
pressure of the cosmic fluid respectively. Note that $\rho_{m}$ is the proper
isotropic density of matter-radiation, $\rho_{\phi}$ denotes the density of
the scalar field and $P_{m}$, $P_{\phi}$ are the corresponding isotropic
pressures. In the context of a FRW metric in Cartesian coordinates
\begin{equation}
ds^{2}=-dt^{2}+a^{2}(t)\frac{1}{(1+\frac{\mathit{K}}{4}\mathbf{x}^{2})^{2}%
}(dx^{2}+dy^{2}+dz^{2}) \label{SF.1}%
\end{equation}
the Einstein's field equations (\ref{EE}), for comoving observers ($U^{\mu
}=\delta_{0}^{\mu}$), provide
\begin{align}
R_{00}  &  =-3\frac{\ddot{a}}{a}\label{SF.2}\\
R_{\mu\nu}  &  =\left[  \frac{\ddot{a}}{a}+2\frac{\dot{a}^{2}+\mathit{K}%
}{a^{2}}\right]  g_{\mu\nu} \label{SF.3a}%
\end{align}%
\begin{equation}
R=g^{\mu\nu}R_{\mu\nu}=6\left[  \frac{\ddot{a}}{a}+\frac{\dot{a}%
^{2}+\mathit{K}}{a^{2}}\right]  \label{SF.3b}%
\end{equation}
where the over-dot denotes derivative with respect to the cosmic time $t$,
$a(t)$ is the scale factor of the universe and $\mathit{K}=0,\pm1$ is the
spatial curvature parameter. Finally, the gravitational field equations boil
down to Freedman's equation
\begin{equation}
H^{2}\equiv\left(  \frac{\dot{a}}{a}\right)  ^{2}=\frac{k}{3}\left(  \rho
_{m}+\rho_{\phi}\right)  -\frac{\mathit{K}}{a^{2}}, \label{frie1}%
\end{equation}
and
\begin{equation}
3H^{2}+2\dot{H}=-k(P_{m}+P_{\phi})-\frac{\mathit{K}}{a^{2}} \label{frie2}%
\end{equation}
where $H(t)\equiv\dot{a}/a$ is the Hubble function. The Bianchi identity
$\bigtriangledown^{\mu}\,{\tilde{T}}_{\mu\nu}=0$ amounts to the following
generalized local conservation law:%

\begin{equation}
\dot{\rho}_{m}+\dot{\rho_{\phi}}+3H(\rho_{m}+P_{m}+\rho_{\phi}+P_{\phi})=0\,.
\label{frie3}%
\end{equation}
Combining eqs. (\ref{frie1}), (\ref{frie2}) and (\ref{frie3}) we obtain
\begin{equation}
\frac{\ddot{a}}{a}=-\frac{k}{6}\,[\rho_{m}+\rho_{\phi}+3\,(P_{m}+P_{\phi})].
\label{2FL}%
\end{equation}
Assuming negligible interaction between matter and the scalar field we have:
\begin{equation}
(\rho_{m},P_{m})\equiv(-T_{0}^{0},T_{i}^{i})\;\;\;\;(\rho_{\phi},P_{\phi
})\equiv(-T_{0}^{0}(\phi),T_{i}^{i}(\phi)). \label{cons}%
\end{equation}
Then eq. (\ref{frie3}) leads to the following independent differential
equations
\begin{equation}
\dot{\rho}_{m}+3H(\rho_{m}+P_{m})=0\, \label{frie4}%
\end{equation}%
\begin{equation}
\dot{\rho}_{\phi}+3H(\rho_{\phi}+P_{\phi})=0\, \label{frie5}%
\end{equation}
and the corresponding equation of state (EoS) parameters are given by
$w_{m}=P_{m}/\rho_{m}$ and $w_{\phi}=P_{\phi}/\rho_{\phi}$. In what follows we
assume a constant $w_{m}$ which implies that $\rho_{m}=\rho_{m0}%
a^{-3(1+w_{m})}$ ($w_{m}=0$ for cold matter and $w_{m}=1/3$ for \ relativistic
matter), where $\rho_{m0}$ is the matter density at the present time.
Generically, some high energy field theories suggest that the dark energy EoS
parameter is a function of cosmic time (see, for instance, \cite{Ellis05}) and
thus
\begin{equation}
\rho_{\phi}(a)=\rho_{\phi0}\;{\exp}\left(  \int_{a}^{1}\frac{3[1+w_{\phi
}(\sigma)]}{\sigma}d\sigma\right)  \label{frie55}%
\end{equation}
where $\rho_{\phi0}$ is the DE density at the current epoch.

\subsection{Scalar field cosmology}

We consider a scalar field in a FRW background which is minimally coupled to
gravity, such that the field satisfies the Cosmological Principle that is,
$\phi$ inherits the symmetries of the metric. This means that the scalar field
depends only on the cosmic time $t$ and consequently $\phi_{,\nu}=\dot{\phi
}\delta_{\nu}^{0}$ where $\dot{\phi}=\frac{d\phi}{dt}$. A scalar field
$\phi(t)$ with a potential $V(\phi)$ is defined by the energy momentum tensor
of the form (for review see \cite{Ame10} and references therein)
\begin{equation}
T_{\mu\nu}(\phi)=-\frac{2}{\sqrt{-g}}\frac{\delta(\sqrt{-g}L_{\phi})}{\delta
g^{\mu\nu}}\label{tensor2}%
\end{equation}
where $L_{\phi}$ is the Lagrangian of the scalar field. Although in the
current analysis we study generically, as much as possible, the problem we
will focus on a scalar field with
\begin{equation}
L_{\phi}=-\frac{1}{2}\varepsilon g^{\mu\nu}\phi_{,\mu}\phi_{,\nu}%
-V(\phi)\label{SF.32}%
\end{equation}
or equivalently
\begin{equation}
L_{\phi}=\frac{1}{2}\varepsilon\dot{\phi}^{2}-V(\phi)\label{Lag1}%
\end{equation}
where
\begin{equation}
\varepsilon=\left\{
\begin{array}
[c]{cc}%
1 & \mbox{Quintessence}\\
-1 & \mbox{Phantom\;.}
\end{array}
\right.
\end{equation}
Therefore, using the second equality of eq.(\ref{cons}), eq.({\ref{tensor2})
and eq.({\ref{Lag1}) the energy density $\rho_{\phi}$ and the pressure
$P_{\phi}$ of the scalar field are given by
\begin{equation}
\rho_{\phi}\equiv-T_{0}^{0}(\phi)=\frac{1}{2}\varepsilon\dot{\phi}^{2}%
+V(\phi)\label{den1}%
\end{equation}
and
\begin{equation}
P_{\phi}\equiv T_{i}^{i}(\phi)=L_{\phi}=\frac{1}{2}\varepsilon\dot{\phi}%
^{2}-V(\phi)\;\;.\label{pres1}%
\end{equation}
Inserting eq.(\ref{den1}) and eq.(\ref{pres1}) into eq.(\ref{frie5}) we derive
the Klein-Gordon equation which describes the time evolution of the scalar
field. This is
\begin{equation}
\ddot{\phi}+\frac{3}{a}\dot{a}\dot{\phi}+\varepsilon V_{,\phi}=0\label{klein1}%
\end{equation}
where $V_{,\phi}=dV/d\phi$. If we use the current functional form of $L_{\phi
}$ then eq.(\ref{frie2}) takes the form:
\begin{equation}
\frac{\ddot{a}}{a}+\frac{1}{2}\left(  \frac{\dot{a}^{2}}{a^{2}}+\frac
{\mathit{K}}{a^{2}}\right)  +\frac{k}{2}\left(  P_{m}+\frac{1}{2}%
\varepsilon\dot{\phi}^{2}-V(\phi)\right)  =0\label{klein2}%
\end{equation}
}}Notice, that for the rest of the paper we use a spatially flat FLRW metric,
namely $\mathit{K}=0.$

The corresponding dark energy EoS parameter is
\begin{equation}
w_{\phi}=\frac{P_{\phi}}{\rho_{\phi}}=\frac{\varepsilon(\dot{\phi}%
^{2}/2)-V(\phi)}{\varepsilon(\dot{\phi}^{2}/2)+V(\phi)}\;\;. \label{pres11}%
\end{equation}
The quintessence ($\varepsilon=1$) cosmological model accommodates a late time
cosmic acceleration in the case of $w_{\phi}<-1/3$ which implies that
$\dot{\phi}^{2}<V(\phi)$. On the other hand, if the kinetic term of the scalar
field is negligible with respect to the potential energy [i.e. $\frac
{\dot{\phi}^{2}}{2}\ll V(\phi)$] then the equation of state parameter is
$w_{\phi}\simeq-1$. In the case of a phantom DE ($\varepsilon=-1$), due to the
negative kinetic term, one has $w_{\phi}<-1$ and $(\dot{\phi}^{2}/2)<V(\phi)$.

The unknown quantities of the problem are $a(t),$ $\phi(t)$ and $V(\phi)$
whereas we have only two independent differential equations available namely
eqs. (\ref{klein1}) and (\ref{klein2}). Therefore in order to solve the system
of differential equations we need to assume a functional form of the scalar
field potential $V(\phi)$. In the literature, due to the unknown nature of DE,
there are many forms of this potential (for a review see \cite{Ame10}) which
describe differently the physical features of the scalar field. In the present
work we use dynamical symmetries of the field equations in order to determine
the unknown potential $V\left(  \phi\right)  $.

\section{Lie B\"{a}cklund symmetries}

\label{LBsym}

In this section we give the basic definitions and properties for the
generalized symmetries. Consider a function $H\left(  x^{i},u^{A},u_{,i}%
^{A},u_{,ij}^{A}...\right)  $ in the space $B_{M}\left\{  x^{i},u^{A}%
,u_{,i}^{A},u_{,ij}^{A},...\right\}  $ where $x^{i}$ are $n$ independent
variables and $u^{A}$ are $m$ dependent variables. The infinitesimal
transformation
\begin{align}
\bar{x}^{i}  &  =x^{i}+\varepsilon\xi^{i}\left(  x^{i},u^{B},u_{,i}%
^{B},u_{,ij}^{B}...\right) \label{LB.01}\\
\bar{u}^{A}  &  =u^{A}+\varepsilon\eta^{A}\left(  x^{i},u^{B},u_{,i}%
^{B},u_{,ij}^{B}...\right)  \label{LB.02}%
\end{align}
with generator
\begin{equation}
X=\xi^{i}\left(  x^{i},u^{B},u_{,i}^{B},u_{,ij}^{B}...\right)  \partial
_{i}+\eta^{A}\left(  x^{i},u^{B},u_{,i}^{B},u_{,ij}^{B}...\right)
\partial_{u} \label{LB.03}%
\end{equation}
is called a Lie B\"{a}cklund symmetry of the differential equation
\begin{equation}
H\left(  x^{i},u^{A},u_{,i}^{A},u_{,ij}^{A}...\right)  =0 \label{LB.03b}%
\end{equation}
if and only if there exist a function $\lambda\left(  x^{i},u,u_{,i}%
,u_{,ij}...\right)  $ such that \cite{Stephani,Bluman}
\begin{equation}
\left[  X,H\right]  =\lambda H~\ ,~\mathrm{mod}H=0. \label{LB.03bA}%
\end{equation}

From the above definition it follows that a Lie B\"{a}cklund symmetry
preserves the set of solutions $u$ of $H\left(  x^{i},u,u_{,i},u_{,ij}%
...\right)  $. In the case where the generator (\ref{LB.03}) of the
infinitesimal transformation (\ref{LB.01}), (\ref{LB.02}) depends only on the
variables $\left\{  x^{i},u^{A}\right\}  $, i.e. $\frac{\partial\xi^{i}%
}{\partial u_{,ij..}^{B}}=\frac{\partial\eta^{A}}{\partial u_{,ij..}^{B}}=0$;
the infinitesimal transformation (\ref{LB.01}), (\ref{LB.02}) is a point
transformation and the generator $X$ is a Lie point symmetry if there exist
$\lambda$ such as condition (\ref{LB.03bA}) holds. That is, the Lie
B\"{a}cklund symmetries are more general and reduce to the Lie point
symmetries when the generator $X$ is independent of the derivatives. In the
following we consider only Lie B\"{a}cklund symmetries.

The operator $D_{i}=\partial_{i}+u_{,i}\partial_{u}+u_{,ij}\partial_{u,_{i}%
}+...$ defines always a Lie B\"{a}cklund symmetry (the trivial one)
\cite{Stephani}. Therefore, if (\ref{LB.03}) is a Lie B\"{a}cklund symmetry of
$H$ then the generator
\[
\bar{X}=X-f^{i}D_{i}=\left(  \xi^{k}-f^{k}\right)  \partial_{k}+\left(
\eta^{A}-f^{k}u_{,k}^{A}\right)  \partial_{u^{A}}+...
\]
is also a Lie B\"{a}cklund symmetry. Since $f^{k}$ is an arbitrary function we
set $f^{k}=\xi^{k}$ and obtain%
\begin{equation}
\bar{X}=\left(  \eta^{A}-\xi^{k}u_{,k}^{A}\right)  \partial_{u^{A}}.
\label{LB.04}%
\end{equation}
The generator (\ref{LB.04}) is the canonical form of the Lie B\"{a}cklund
symmetry (\ref{LB.03}). Furthermore we can always absorb the term $\xi
^{k}u_{k}$ inside the $\eta$ and conclude that $\bar{X}=Z^{A}\left(
x^{i},u^{B},u_{,i}^{B},u_{,ij}^{B}...\right)  \partial_{u^{A}}$ is the
generator of a Lie B\"{a}cklund symmetry. A special class of Lie B\"{a}cklund
symmetries are the contact symmetries defined by the requirement that the
generator depends only on the first derivatives $u_{,i}$, i.e. it has the
general canonical form%
\begin{equation}
X_{C}=Z^{A}\left(  x^{i},u^{B},u_{,i}^{B}\right)  \partial_{u^{A}}.
\label{LB.04a}%
\end{equation}

\subsection{Dynamical Noether symmetries}

\label{DynSym}

Suppose that the dynamical system (\ref{LB.03b}) follows from a variational
principle, that is, equations (\ref{LB.03b}) are the Euler-Lagrange equations
for a Lagrangian function $L\left(  x^{i},u^{A},u_{,i}^{A}...\right)  .~$The
vector field $\bar{X}=Z^{i}\left(  t,q^{k},\dot{q}^{k}\right)  \partial
_{q^{i}}~$where $\dot{q}^{i}=\frac{dq^{i}}{dt}$ and $Z^{i}\left(  t,q^{k}%
,\dot{q}^{k}\right)  $ is linear in $\dot{q}^{k}$ is called a dynamical
(contact) Noether symmetry of the Lagrangian $L\left(  t,q^{i},\dot{q}%
^{i}\right)  $ if there exist a function $f\left(  t,q^{i}\right)  $ such that
the following condition holds \cite{Sarlet}%
\begin{equation}
\bar{X}^{\left[  1\right]  }L=\dot{f}\left(  t,q^{i}\right)  \label{LB.05}%
\end{equation}
where $\bar{X}^{\left[  1\right]  }$ is the first prolongation of $\bar{X}$,
i.e. $\bar{X}^{\left[  1\right]  }=X+\dot{Z}^{i}\partial_{q^{i}}$.

If $\bar{X}$ is a dynamical Noether symmetry of $L\left(  t,q^{i},\dot{q}%
^{i}\right)  $, then the quantity \cite{Sarlet,Crampin}
\begin{equation}
I=Z^{i}\left(  t,q^{k},\dot{q}^{k}\right)  \frac{\partial L}{\partial\dot
{q}^{i}}-f\left(  t,q^{i}\right)  \label{LB.05A}%
\end{equation}
is a first integral of Lagrange equations and it is called a (contact) Noether
Integral. When $Z^{i}\left(  t,q^{k},\dot{q}^{k}\right)  =Z^{i}\left(
t,q^{k}\right)  $ the integral $I\left(  t,q^{k},\dot{q}^{k}\right)  $ it is
linear in the momentum and in that case the Noether symmetry~$\bar{X}~$ it is
a~Noether point symmetry.

Consider a particle moving in an $n$ dimensional Riemannian space with metric
$g_{ij}\left(  q^{k}\right)  $ under the action of the potential $V\left(
q^{k}\right)  .$ The Lagrangian of the system is
\begin{equation}
L\left(  q^{k},\dot{q}^{k}\right)  =\frac{1}{2}g_{ij}\dot{q}^{i}\dot{q}%
^{j}-V\left(  q^{k}\right)  . \label{LB.06}%
\end{equation}
Let $\bar{X}=K_{j}^{i}\left(  t,q^{k}\right)  \dot{q}^{i}\partial_{i}$ be the
generator of a contact Lie B\"{a}cklund symmetry of (\ref{LB.06}). In
\cite{Kalotas} it has been shown that in this case the dynamical symmetry
condition (\ref{LB.05}) is equivalent to the the following conditions
\begin{equation}
K_{\left(  ij;k\right)  }=0~ \label{LB.07}%
\end{equation}%
\begin{equation}
K_{ij,t}=0~~,~f_{,t}=0 \label{LB.08}%
\end{equation}%
\begin{equation}
K^{ij}V_{j}+f_{,i}=0. \label{LB.09}%
\end{equation}
where $";"$ denotes covariant derivative with respect to the connection
coefficients of the metric $g_{ij}$.

From (\ref{LB.08}) it follows that $K_{j}^{i}=K_{j}^{i}\left(  q^{k}\right)  $
and $f=f\left(  q^{k}\right)  $. Furthermore, condition (\ref{LB.07}) means
that the second rank tensor $K_{j}^{i}\left(  q^{k}\right)  $ is a Killing
tensor of the metric $g_{ij}~$. Finally (\ref{LB.09}) is a constraint relating
the potential with the Killing tensor $K^{ij}$ and the Noether function $f.$
The use of dynamical Noether symmetries provides first integrals which can be
used to reduce the order of the dynamical system and possibly lead to analytic solutions.

The application of the Noether point symmetries in scalar field cosmology has
been studied in \cite{Basilakos}. In this work we would to extend the analysis
to the case of dynamical (contact) Noether symmetries. In the following
section we use the symmetry condition (\ref{LB.09}) in order to identify the
potential(s) of the scalar field in scalar field cosmology for which the field
equations admit contact Noether symmetries. Subsequently we use the conserved
currents of these symmetries to determine analytic solutions of the resulting
scalar field equations.

\section{Dynamical Noether symmetries in scalar field cosmology}

\label{DynNSF}

Consider a dynamical system which consists of a minimally coupled scalar field
and dust (DM component) in the flat FRW background (\ref{SF.1}). The
gravitational field equations are the Euler Lagrange equations of the
Lagrangian
\begin{equation}
L\left(  a,\phi,\dot{a},\dot{\phi}\right)  =-3a\dot{a}^{2}+\frac{\varepsilon
}{2}a^{3}\dot{\phi}^{2}-a^{3}V\left(  \phi\right)  \label{CS.05}%
\end{equation}
with Hamiltonian
\begin{equation}
E=-3a\dot{a}^{2}+\frac{\varepsilon}{2}a^{3}\dot{\phi}^{2}+a^{3}V\left(
\phi\right)  \label{CS.06}%
\end{equation}
where $\rho_{m}=\left\vert E\right\vert a^{-3}~$and $E$ is a
constant\footnote{Where we have set $k=1.$}, hence the today value of
$\rho_{m}$ is $\rho_{m0}=\left\vert E\right\vert $~and $\rho_{m0}=3\omega
_{m0}$ where $\omega_{m0}=\Omega_{m0}H_{0}^{2}$.

Using the coordinate transformation $\left(  a,\phi\right)  \rightarrow\left(
r,\theta\right)  $ defined by%
\begin{equation}
r=\sqrt{\frac{8}{3}}a^{\frac{3}{2}}~,~\theta=\sqrt{\frac{3\varepsilon}{8}}%
\phi\label{CS.07}%
\end{equation}
the Lagrangian (\ref{CS.05}) becomes
\begin{equation}
L\left(  r,\theta,\dot{r},\dot{\theta}\right)  =-\frac{1}{2}\dot{r}^{2}%
+\frac{1}{2}r^{2}\dot{\theta}^{2}-r^{2}V\left(  \theta\right)  . \label{CS.08}%
\end{equation}

The potential of the scalar field is yet unspecified. In order to select a
potential we make the geometric assumption that Lagrange equations admit a
dynamical (contact) Noether symmetry. As it has been shown in the last section
this requirement is equivalent to condition (\ref{LB.05}). From Lagrangian
(\ref{CS.08}) we infer that the kinetic metric is the 2d minisuperspace with
line element%
\begin{equation}
ds^{2}=-dr^{2}+r^{2}d\theta^{2} \label{CS.09}%
\end{equation}
whereas the effective potential is $V_{eff}\left(  r,\theta\right)
=r^{2}V\left(  \theta\right)  $.

Dynamical Noether symmetries require the knowledge of Killing tensors of
valence two in the minisuperspace (\ref{CS.09}). This space is 2d flat hence
the Killing tensors $K_{ij}$ form a six dimensional space and all are
constructed from the symmetrized products of the KVs (see e.g. \cite{Chanu}).
In Cartesian coordinates\footnote{The coordinate transformation is
\[
r^{2}=\left(  x^{2}-y^{2}\right)  ~,~\theta=\arctan h\left(  \frac{y}%
{x}\right)
\]
} the generic form of a second rank Killing tensor in (\ref{CS.09}) is
\cite{Chanu}
\[
K_{ij}=%
\begin{pmatrix}
c_{1}y^{2}+2c_{2}y+c_{3} & c_{6}-c_{1}yx-c_{2}x-c_{4}y\\
c_{6}-c_{1}yx-c_{2}x-c_{4}y & c_{1}x^{2}+2c_{4}x+c_{5}%
\end{pmatrix}
.
\]

We apply condition (\ref{LB.09}) taking into account the above result. For
arbitrary potential $V\left(  \theta\right)  $ \ the Lagrangian (\ref{CS.08})
admits only the trivial contact symmetry $X_{H}=g_{j}^{i}\dot{x}^{j}%
\partial_{i}$ where $g_{ij}$ is the two dimensional kinetic metric
(\ref{CS.09}). The corresponding Noether integral of this symmetry is the
Hamiltonian constraint.

In order to find extra dynamical Noether symmetries we must consider special
forms for the potential~$V\left(  \theta\right)  $. \ A detailed analysis
gives the following results\footnote{In appendix \ref{appendix1} we give the
complete classification of the potentials for which Lagrangian (\ref{CS.08})
admits contact Noether symmetries.}:

\begin{itemize}
\item For the potential
\begin{equation}
V\left(  \theta\right)  =c_{1}+c_{2}\cosh^{2}\theta\label{Pot.1}%
\end{equation}
Lagrangian (\ref{CS.08}) admits the additional dynamical symmetry%
\begin{equation}
X=-\left(  \cosh^{2}\theta\dot{r}+\frac{1}{2}r\sinh\left(  2\theta\right)
\dot{\theta}\right)  \partial_{r}+\frac{1}{r}\left(  \frac{1}{2}\sinh\left(
2\theta\right)  \dot{r}+r\sinh\theta~\dot{\theta}\right)  \partial_{\dot{r}}%
\end{equation}
with corresponding Noether Integral%
\begin{equation}
I_{1}=\left(  \cosh\theta\dot{r}+r\sinh\theta~\dot{\theta}\right)  ^{2}%
-2r^{2}\left(  c_{1}+c_{2}\right)  \cosh^{2}\theta.
\end{equation}
The model with potential (\ref{Pot.1}) is the well known UDM model
\cite{Basilakos,BasilLukes}. If \thinspace$c_{2}=3c_{1}$ the potential
(\ref{Pot.1}) admits one more \ dynamical symmetry%
\begin{equation}
X=-r^{2}\sinh\theta~\dot{\theta}\partial_{r}+\left(  \sinh\theta\dot
{r}+2r\cosh\theta~\dot{\theta}\right)  \partial_{\theta} \label{Pot2l}%
\end{equation}
with corresponding Noether Integral%
\begin{equation}
\bar{I}_{1}=\left(  r^{2}\sinh\theta~\dot{r}\dot{\theta}+r^{3}\cosh\theta
~\dot{\theta}^{2}\right)  +2c_{1}r^{3}\cosh\theta\sinh^{2}\theta
\end{equation}

\item For the potential%
\begin{equation}
V\left(  \theta\right)  =c_{1}\left(  1+3\cosh^{2}\theta\right)  +c_{2}\left(
3\cosh\theta+\cosh^{3}\theta\right)  \label{Pot.2}%
\end{equation}
Lagrangian (\ref{CS.08}) admits the dynamical symmetry (\ref{Pot2l}) with
corresponding Noether Integral%
\begin{equation}
I_{2}=\left(  r^{2}\sinh\theta~\dot{r}\dot{\theta}+r^{3}\cosh\theta
~\dot{\theta}^{2}\right)  +r^{3}\sinh^{2}\theta\left(  2c_{1}\cosh\theta
+c_{2}\left(  1+\cosh^{2}\theta\right)  \right)
\end{equation}
In case $c_{1}=c_{2}$ the potential becomes%
\[
V\left(  \theta\right)  =c_{1}\left(  1+\cosh\theta\right)  ^{3}.
\]
which is the model of Sahni et.al \cite{Sahni} (for $p=3$).
\end{itemize}

\section{Analytic solutions}

\label{AnS}

It is straightforward to see that the dynamical Noether integrals $I_{1}%
,I_{2}$ are in involution and independent of the Hamiltonian $H$,
i.e.$\left\{  I,H\right\}  =0$, hence the dynamical systems we have found are
Liouville integrable. In this section we apply the extra integrals to reduce
the order of the dynamical system and if it is feasible to find an exact
solution. In the following the constants $c_{1},c_{2}$ are assumed to be
positive. We would like to mention that analytical solutions in the context of
an hyperbolic type potential can be also found \cite{BarrowR}.

%\subsection{Model (\ref{Pot.1})}

\subsection{The Unified Dark Matter Model}

The quintessence UDM cosmological model has been studied both analytically and
statistically in \cite{BasilLukes} (see also
\cite{Ber07,Gorini04,Gorini05,Basilakos,capPhantom,BasilLukes}). In the latter
papers it has been found that the quintessence UDM scalar field model is in a
fair agreement with that of the $\Lambda$-cosmology\footnote{Recall that for
the $\Lambda$-cosmology the exact solution of the scale factor is%
\[
a_{\Lambda}\left(  t\right)  =\left(  \frac{\Omega_{m0}}{1-\Omega_{m0}%
}\right)  ^{\frac{1}{3}}\sinh^{\frac{2}{3}}\left(  \omega_{1}t\right)  \;.
\]
The Hubble function is written as
\[
H_{\Lambda}(t)=\frac{2}{3}\omega_{1}\coth(\omega_{1}t)
\]
where $\omega_{1}=\frac{3H_{0}(1-\Omega_{m0})^{1/2}}{2}$.} at expansion and at
perturbation levels, although there are some differences between the two
models. In the current work we solve analytically the UDM dynamical problem by
treating dark energy simultaneously either as quintessence or phantom.
Moreover we have provided, for a first time (to our knowledge), the dynamical
Noether symmetries of the UDM model. Evidently, the combination of the results
published by Basilakos \& Lukes \cite{BasilLukes} and Basilakos et al.
\cite{Basilakos} with the current article provide a complete investigation of
the UDM scalar field model.

Using the notations of Bertacca et al. \cite{Ber07}, the real constants in Eq.
(\ref{Pot.1}) are chosen to obey $c_{1}=c_{2}>0$. It is interesting to mention
that the potential (\ref{Pot.1}) has one minimum at $\phi=0$, which reads
\begin{equation}
V_{min}=V(0)=c_{1}+c_{2}.
\end{equation}
Lastly, as long as the scalar field is taking negative and large values the
UDM model has the attractive feature due to $V(\theta)\propto e^{-2\theta}$
\cite{Sahni}.

\subsubsection{UDM: Quintessence}

\label{modelUDMq} Inserting in Eq.(\ref{CS.05}) Eq.(\ref{Pot.1}) and
$\varepsilon=1$ the Lagrangian of the field equations becomes
\begin{equation}
L\left(  a,\phi,\dot{a},\dot{\phi}\right)  =-3a\dot{a}^{2}+\frac{1}{2}%
a^{3}\dot{\phi}^{2}-a^{3}\left(  c_{Q1}+c_{Q2}\cosh^{2}\sqrt{\frac{3}{8}}%
\phi\right)  \label{ud.1}%
\end{equation}
where the index $Q$ denotes the quintessence model. Under the coordinate
transformation
\[
a^{3}=\frac{3}{8}\left(  x^{2}-y^{2}~\right)  ,~\phi=\sqrt{\frac{8}{3}}\arctan
h\left(  \frac{y}{x}\right)
\]
the Lagrangian becomes%
\begin{equation}
L=\frac{1}{2}\left(  -\dot{x}^{2}+\dot{y}^{2}\right)  -\frac{1}{2}\left(
\omega_{1}^{2}x^{2}-\omega_{2}^{2}y^{2}\right)
\end{equation}
which is the Lagrangian of the 2d unharmonic hyperbolic oscillator. The field
equations are the Hamiltonian constraint%
\begin{equation}
\frac{1}{2}\left(  -p_{x}^{2}+p_{y}^{2}\right)  +\frac{1}{2}\left(  \omega
_{1}^{2}x^{2}-\omega_{2}^{2}y^{2}\right)  =E\label{ud.2}%
\end{equation}
and Hamilton's equations of (\ref{ud.2}). Furthermore $\omega_{1}^{2}=\frac
{3}{4}\left(  c_{Q1}+c_{Q2}\right)  $, $\omega_{2}^{2}=\frac{3}{4}c_{Q1}$ are
the oscillators' \textquotedblleft frequencies\textquotedblright\ with units
of inverse of time ($\omega_{1,2}\propto H_{0}$) and $p_{x},p_{y}$ are the
components of the momentum. Since $c_{Q1}=c_{Q2}$ we simply derive $\omega
_{1}=\sqrt{2}\omega_{2}$.

The solution of the field equations is%
\begin{align}
x\left(  t\right)   &  =x_{0}\sinh\left(  \omega_{1}t+\theta_{1}\right)  \\
y\left(  t\right)   &  =y_{0}\sinh\left(  \omega_{2}t+\theta_{2}\right)  .
\end{align}
The Hamiltonian constraint (\ref{ud.2}) gives $E=\frac{1}{2}\left(  \omega
_{2}^{2}y_{0}^{2}-\omega_{1}^{2}x_{0}^{2}\right)  $. Moreover close to the
singularity $a\left(  t\rightarrow0^{+}\right)  \rightarrow0^{+}$ we have the
constraint
\begin{equation}
\theta_{2}=\pm\arcsin h\left(  \frac{x_{0}}{y_{0}}\sinh\theta_{1}\right)
.\label{ud.3}%
\end{equation}
Without loosing the generality we set $x_{0}=y_{0}$. In that case from the
Hamiltonian constraint we have$~\left\vert E\right\vert =\frac{1}{4}x_{0}%
^{2}\omega_{1}^{2}$. This gives $x_{0}^{2}=\frac{8}{3}\frac{\left\vert
E\right\vert }{c_{Q1}}$ and the solution of the scale factor becomes%
\begin{equation}
a^{3}\left(  t\right)  =\frac{9}{2}\frac{\omega_{m0}}{\omega_{1}^{2}}\left[
\sinh^{2}\left(  \omega_{1}t+\theta_{1}\right)  -\sinh^{2}\left(  \frac
{\sqrt{2}}{2}\omega_{1}t+\varepsilon_{\theta}\theta_{1}\right)  \right]
\label{ud.4}%
\end{equation}
where $\varepsilon_{\theta}=\pm1$. Equation (\ref{ud.4}) can be written
\begin{equation}
a\left(  t\right)  =a_{-}\left(  t\right)  \sinh^{\frac{2}{3}}\left(
\omega_{1}t+\theta_{1}\right)  \label{ud.4a}%
\end{equation}
where
\begin{equation}
a_{-}^{3}\left(  t\right)  =\frac{9}{2}\frac{\omega_{m0}}{\omega_{1}^{2}%
}\left[  1-\left(  \frac{\sinh\left(  \frac{\sqrt{2}}{2}\omega_{1}%
t+\varepsilon_{\theta}\theta_{1}\right)  }{\sinh\left(  \omega_{1}t+\theta
_{1}\right)  }\right)  ^{2}\right]
\end{equation}
with limit $a_{-}\left(  t\right)  |_{\omega_{1}t+\theta_{1}>>1}=\frac{9}%
{2}\frac{\omega_{m0}}{\omega_{1}^{2}}=const$. Therefore for the late time,
$\omega_{1}t+\theta_{1}\simeq\omega_{1}t$ the scale factor (\ref{ud.4a})
becomes the $\Lambda-$cosmology. Furthermore for the late time, $\omega
_{1}t+\theta_{1}\simeq\omega_{1}t$ the Hubble function for the scale factor
(\ref{ud.4a}) is%
\begin{equation}
H\left(  t\right)  =H_{\Lambda}\left(  t\right)  +\frac{\dot{a}_{-}\left(
t\right)  }{a_{-}\left(  t\right)  }%
\end{equation}
where $H_{\Lambda}\left(  t\right)  $ is the Hubble function of the $\Lambda$
cosmology$.$

\begin{figure}[ptb]
\includegraphics[height=8cm]{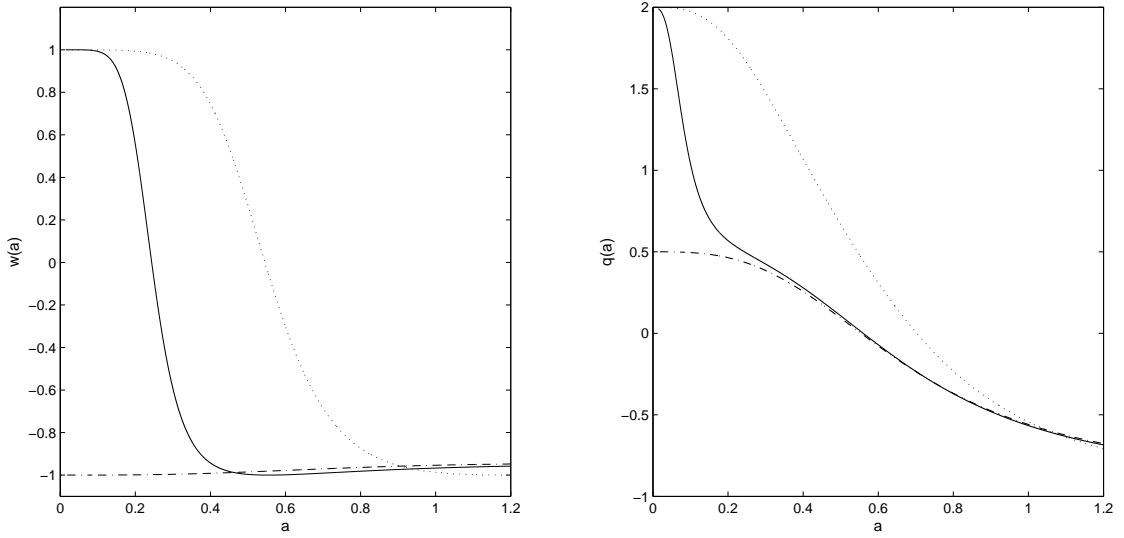}
%\mbox{\epsfxsize=14.2cm \epsffile{NumericUDMQ.eps}}
\caption{Left: Equation of state $w_{\phi}\left(  a\right)  $ evolution for
scalar field with scale factor (\ref{ud.4}). Right: Deceleration parameter
$q\left(  a\right)  $ for scale factor (\ref{ud.4}). We use $\left(
\omega_{m0},c_{Q1}\right)  =\left(  0.12,0.36\right)  \times10^{4}$,
$\varepsilon_{\theta}=-1$ $;$ solid line is for $\theta_{1}=0.01$, the dot
line is for $\theta_{1}=0.1$ and the dash dot line is for $\theta_{1}=0.$}%
\label{numer1}%
\end{figure}

In$~$Fig. \ref{numer1} we present the evolution of the equation of state
parameter of the scalar field and the evolution of the deceleration parameter
of the scale factor (\ref{ud.4}). We observe that for values of $\theta_{1}%
\in\left(  0,0.1\right)  $ the equation of state parameter has values
$w_{\phi}\in\left[  -1,1\right]  $ and reaches the value $-1$ for large scale
factor, however for $\theta_{1}=0~$the equation of state parameter is
$w_{\phi}\in\lbrack-1,-0.95)$.

\subsubsection{UDM: Phantom}

For $\varepsilon=-1$ and the potential (\ref{Pot.1}) the Lagrangian of the
field equations becomes%
\begin{equation}
L\left(  a,\phi,\dot{a},\dot{\phi}\right)  =-3a\dot{a}^{2}-\frac{1}{2}%
a^{3}\dot{\phi}^{2}-a^{3}\left(  c_{p1}+c_{p2}\cos^{2}\sqrt{\frac{3}{8}}%
\phi\right)  \label{pud.1}%
\end{equation}
where the index $P$ denotes the phantom model. Under the coordinate
transformation%
\[
a^{3}=\frac{3}{8}\left(  \bar{x}^{2}+\bar{y}^{2}~\right)  ,~\phi=\sqrt
{\frac{8}{3}}\arctan\left(  \frac{\bar{y}}{\bar{x}}\right)
\]
the field equations are the Hamiltonian constraint%
\begin{equation}
-\frac{1}{2}\left(  p_{\bar{x}}^{2}+p_{\bar{y}}^{2}\right)  +\frac{1}%
{2}\left(  \bar{\omega}_{1}^{2}\bar{x}^{2}+\bar{\omega}_{2}^{2}\bar{y}%
^{2}\right)  =E. \label{pud.2}%
\end{equation}
and the Hamilton's equations of (\ref{pud.2}). This dynamical system is the 2d
unharmonic hyperbolic oscillator where $p_{\bar{x}},p_{\bar{y}}$ are the
components of the momentum and $\bar{\omega}_{1}^{2}=\frac{3}{4}\left(
c_{p1}+c_{p2}\right)  $, $\bar{\omega}_{2}^{2}=\frac{3}{4}c_{p_{1}}.~$ The
solution of the system is%
\begin{align*}
\bar{x}\left(  t\right)   &  =\bar{x}_{0}\sinh\left(  \bar{\omega}_{1}%
t+\bar{\theta}_{1}\right) \\
\bar{y}\left(  t\right)   &  =\bar{y}_{0}\sinh\left(  \bar{\omega}_{2}%
t+\bar{\theta}_{2}\right)
\end{align*}
for which the Hamiltonian constraint (\ref{pud.2}) gives $E=-\frac{1}%
{2}\left(  \bar{x}_{0}^{2}\bar{\omega}_{1}^{2}+\bar{y}_{0}^{2}\bar{\omega}%
_{2}^{2}\right)  $. Prior to the singularity we have $a\left(  t\rightarrow
0^{+}\right)  \rightarrow0^{+}$ implying $\bar{\theta}_{1}=\bar{\theta}_{2}=0$.

In order to reduce the number of the free parameters we make again the ansazt
$c_{p1}=c_{p2}$ and $\bar{x}_{0}=\bar{y}_{0}$. Then from the Hamiltonian
constraint we have~$\bar{x}_{0}^{2}=\frac{8}{9}\frac{\left\vert E\right\vert
}{c_{p1}}$ and the analytic solution of the scale factor is%
\begin{equation}
a^{3}\left(  t\right)  =\frac{3}{2}\frac{\omega_{m0}}{\bar{\omega}_{1}^{2}%
}\left[  \sinh^{2}\left(  \bar{\omega}_{1}t\right)  +\sinh^{2}\left(
\frac{\sqrt{2}}{2}\bar{\omega}_{1}t\right)  \right]  .\label{pud.222}%
\end{equation}

However, the scalar factor (\ref{pud.222}) can be written in the following
form%
\begin{equation}
a\left(  t\right)  =a_{+}\left(  t\right)  \sinh^{\frac{2}{3}}\left(
\bar{\omega}_{1}t\right)  \label{pud.222a}%
\end{equation}
where%
\begin{equation}
a_{+}^{3}\left(  t\right)  =\frac{3}{2}\frac{\omega_{m0}}{\bar{\omega}_{1}%
^{2}}\left[  1+\left(  \frac{\sinh\left(  \frac{\sqrt{2}}{2}\bar{\omega}%
_{1}t\right)  }{\sinh\left(  \bar{\omega}_{1}t\right)  }\right)  ^{2}\right]
\end{equation}
with limit $a_{+}^{3}\left(  t\right)  |_{\omega_{1}t>>1}=\frac{3}{2}%
\frac{\omega_{m0}}{\bar{\omega}_{1}^{2}}$, hence in the late time the scale
factor (\ref{pud.222a}) is that of the $\Lambda-$cosmology. \ Therefore, for
the Hubble function holds%
\begin{equation}
H\left(  t\right)  =H_{\Lambda}\left(  t\right)  +\frac{\dot{a}_{+}\left(
t\right)  }{a_{+}\left(  t\right)  }.
\end{equation}

%Obviously, from the above analysis it becomes clear that
%for the UDM scalar field the solution of the scale
%factor at the late time can be written in the form $a_{UDM}=a_{\Lambda}\left(
%t\right)  a_{\mp}\left(  t\right)  ~$where~$a_{\Lambda}\left(  t\right)  $ is
%the scale factor of the $\Lambda-$cosmology, $a_{\mp}\left(  t\right)  $ is
%defined as follows
%\begin{equation}
%a_{\mp}\left(  t\right)  =a_{0}\left[  1-\varepsilon\left(  \frac{\sinh\left(
%\frac{\sqrt{2}}{2}\bar{\omega}_{1}t\right)  }{\sinh\left(  \bar{\omega}%
%_{1}t\right)  }\right)  ^{2}\right]  ^{\frac{1}{3}}%
%\end{equation}
%where $\varepsilon$ takes the value $\varepsilon=1$ quintessence scalar field
%and $\varepsilon=-1$ for phantom scalar field.

\subsection{The new hyperbolic model}

In the following we use the second integral $I_{2}$ of the hyperbolic
potential (\ref{Pot.2}) in order to reduce the order of the dynamical system.

\subsubsection{Quintessence, $\varepsilon=1$}

\label{quintsf}

For the potential (\ref{Pot.2}) and $\varepsilon=1,$ we apply the coordinate
transformation (hyper-parabolic coordinates)%
\[
a^{3}=\frac{3}{32}\left(  u^{2}-v^{2}\right)  ^{2}~\ ,~\phi=-\sqrt{\frac{8}%
{3}}\arctan h\left(  \frac{2uv}{u^{2}+v^{2}}\right)
\]
and the Lagrangian (\ref{CS.05}) of the field equations becomes%
\begin{equation}
L\left(  u,v,\dot{u},\dot{v}\right)  =\frac{\left(  u^{2}-v^{2}\right)  }%
{2}\left(  -\dot{u}^{2}+\dot{v}^{2}\right)  -\frac{V_{1}u^{6}-V_{2}v^{6}%
}{u^{2}-v^{2}} \label{ans.01}%
\end{equation}
whereas the Hamiltonian (\ref{CS.06}) is%
\begin{equation}
\left(  u^{2}-v^{2}\right)  ^{-1}\left[  \frac{1}{2}\left(  -p_{u}^{2}%
+p_{v}^{2}\right)  +\frac{V_{1}}{6}u^{6}-\frac{V_{2}}{6}v^{6}\right]  =E
\label{ans.02}%
\end{equation}
where $p_{u},p_{v}$ are momenta and $V_{1}=\frac{9}{4}\left(  c_{Q1}%
+c_{Q2}\right)  ,$ $V_{2}=\frac{9}{4}\left(  c_{Q1}-c_{Q2}\right)  $.

Einstein field equations are the Hamiltonian constraint (\ref{ans.02}) and
Hamilton's equations%
\[
\left(  u^{2}-v^{2}\right)  \dot{u}=-p_{u}~~~~~,~~~\left(  u^{2}-v^{2}\right)
\dot{v}=p_{v}%
\]%
\[
\dot{p}_{u}=\frac{2Eu-V_{1}u^{5}}{\left(  u^{2}-v^{2}\right)  }~~,~~~\ \dot
{p}_{v}=-\frac{2Ev-V_{2}v^{5}}{\left(  u^{2}-v^{2}\right)  }.
\]
In order to solve the system of equations we prefer to work with the Hamilton
Jacobi equation. Hence from (\ref{ans.02}) we have
\begin{equation}
\left(  u^{2}-v^{2}\right)  ^{-1}\left[  \frac{1}{2}\left(  -\left(
\frac{\partial S}{\partial u}\right)  ^{2}+\left(  \frac{\partial S}{\partial
v}\right)  ^{2}\right)  +\frac{V_{1}}{6}u^{6}-\frac{V_{2}}{6}v^{6}\right]
-E\frac{\partial S}{\partial t}=0 \label{ans.03}%
\end{equation}
where $S=S\left(  u,v,t\right)  $ is the Hamiltonian~and $p_{u}=\frac{\partial
S}{\partial u}~,~p_{v}=\frac{\partial S}{\partial v}$. \ \ It is easy to see
that (\ref{ans.03}) is separable, hence the solution is%
\[
S\left(  t,u,v\right)  =-\frac{\sqrt{3}}{3}\int\sqrt{V_{1}u^{6}+6\left\vert
E\right\vert u^{2}+\Phi_{0}}du\pm\frac{\sqrt{3}}{3}\int\sqrt{V_{2}%
v^{6}+6\left\vert E\right\vert v^{2}+\Phi_{0}}dv+t
\]
where $\Phi_{0}$ is an integration constant~ $\left(  \Phi_{0}~\varpropto
~I_{2}\right)  .${\LARGE \ }

Using the Hamiltonian function we find that the reduced system is
\begin{align}
\left(  u^{2}-v^{2}\right)  \dot{u}  &  =\frac{\sqrt{3}}{3}\sqrt{V_{1}%
u^{6}+6\left\vert E\right\vert u^{2}+\Phi_{0}}\label{St.1}\\
~\left(  u^{2}-v^{2}\right)  \dot{v}  &  =\pm\frac{\sqrt{3}}{3}\sqrt
{V_{2}v^{6}+6\left\vert E\right\vert v^{2}+\Phi_{0}}. \label{St.2}%
\end{align}

From the singularity condition $a\left(  t\rightarrow0\right)  =0$ we have
that $\left\vert u\right\vert _{t\rightarrow0}=\left\vert v\right\vert
_{t\rightarrow0}.$However if in $t\rightarrow0^{+}$ we consider $u>v>1~,$ then
from (\ref{St.1}),(\ref{St.2}) we have that $\dot{u}>\dot{v}$, when
$V_{1}>V_{2}$ hence in late time hold that $u^{2}>>v^{2},$ then the system
(\ref{St.1}), (\ref{St.2}) becomes%
\[
\dot{u}=\mu_{1}u~,~\dot{v}=\pm\mu_{2}\frac{v^{3}}{u^{2}}%
\]
where $\mu_{1,2}=\frac{\sqrt{3}}{3}\sqrt{V_{1,2}}$; hence the solution of the
scale factor is%
\begin{equation}
a^{\frac{3}{2}}\left(  t\right)  =a_{1}e^{2\mu_{1}t}\mp\frac{a_{1}^{2}\mu
_{1}^{2}}{\mu_{2}e^{-2\mu_{1}t}+a_{1}^{2}a_{2}\mu_{1}} \label{St.3}%
\end{equation}
when $a_{2}=0$ the scale factor (\ref{St.3}) becomes
\begin{equation}
a^{\frac{3}{2}}\left(  t\right)  =\left(  a_{1}\mp\frac{a_{1}^{2}\mu_{1}^{2}%
}{\mu_{2}}\right)  e^{2\mu_{1}t} \label{St.3a}%
\end{equation}
which is the deSitter solution.

\begin{figure}[ptb]
\includegraphics[height=8cm]{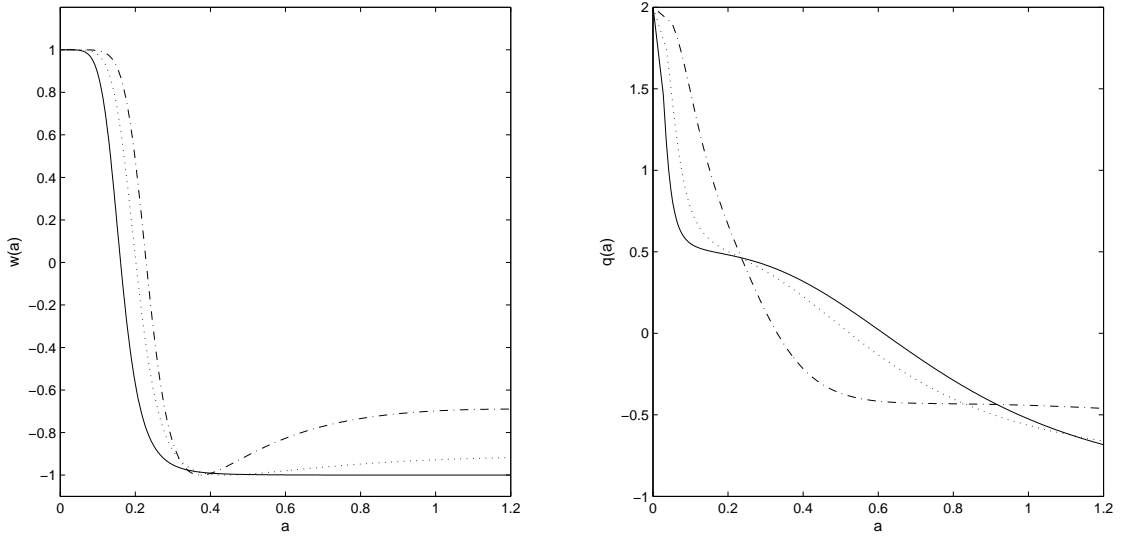}
%\mbox{\epsfxsize=14.2cm \epsffile{numericmodel2q.eps}}
\caption{In the left fig. we give the equation of state $w_{\phi}\left(
a\right)  $ evolution of the scalar field with Lagrangian (\ref{ans.01}) and
in the right fig. we give the deceleration parameter $q\left(  a\right)  $ of
the scale factor with Lagrangian (\ref{ans.01}) where in (\ref{St.2}) we
considered the sing minus. For the numeric solution we use $\left(
u,v\right)  _{t\rightarrow0}=\left(  0.01,0.0998\right)  ,~$ $\left(
\rho_{m0},c_{1},c_{2}\right)  =\left(  0.33,1,0.8\right)  \times10^{4};$ solid
line is for $\Phi_{0}=0$, the dot line is for $\Phi_{0}=10^{3}$ and the dash
dot line is for $\Phi_{0}=10^{4}.$}%
\label{numer2}%
\end{figure}

In$~$Fig. \ref{numer2} $\ $we give the evolution of the equation of state
parameter of the scalar field and the evolution of the deceleration parameter
of the model with Lagrangian (\ref{ans.01}) where in (\ref{St.2}) we
considered the minus. We observe that for the equation of state parameter
$w_{\phi}$ holds $w_{\phi}\left(  a\rightarrow1\right)  =-1$ provided that the
integration constant $\Phi_{0}$ satisfies the condition $\log\Phi_{0}\leq1$.

\paragraph{Exact solution}

The dynamical system (\ref{St.1}), (\ref{St.2}) is a non linear 2d system of
first order ODEs. In order to solve this system analytically we consider the
"conformal" transformation~$\ dt=\sqrt{3}\left(  u^{2}-v^{2}\right)  d\tau
~,~$i.e. $dt\cong a^{\frac{3}{2}}d\tau$ which transforms the dynamical system
to%
\begin{align}
u^{\prime}  &  =\sqrt{V_{1}u^{6}+6\left\vert E\right\vert u^{2}+\Phi_{0}%
}\label{St3}\\
v^{\prime}  &  =\sqrt{V_{2}v^{6}+6\left\vert E\right\vert v^{2}+\Phi_{0}}
\label{St4}%
\end{align}
where $u^{\prime}=\frac{du}{d\tau}$. \ From (\ref{St3}),(\ref{St4}) follows%
\begin{align}
\int\frac{du}{\sqrt{V_{1}u^{6}+6\left\vert E\right\vert u^{2}+\Phi_{0}}}  &
=\int d\tau~~~\label{St5}\\
~\int\frac{dv}{\sqrt{V_{2}v^{6}+6\left\vert E\right\vert v^{2}+\Phi_{0}}}  &
=\int d\tau. \label{St5a}%
\end{align}

The solution of (\ref{St5}), (\ref{St5a}) can be written in terms of elliptic
functions. \ In order to simplify the integrals (\ref{St5}), (\ref{St5a})
\ and obtain an explicit solution we select special values for\ the constants.
For $\Phi_{0}=0$ and $E=0$ the solution of the system (\ref{St5}),(\ref{St5a})
gives the deSitter universe (\ref{St.3a}).

However when $\Phi_{0}=0$ and $E\neq0$ the solution of the system
(\ref{St5}),(\ref{St5a}) is%
\begin{align*}
u\left(  \tau\right)   &  =\frac{2\exp\left(  C_{1}\left(  \tau-\tau
_{0}\right)  \right)  }{\sqrt{\exp\left(  \pm4\sqrt{6\left\vert E\right\vert
}\left(  \tau-\tau_{0}\right)  \right)  -4C_{1}^{2}}}\\
v\left(  \tau\right)   &  =\frac{2\exp\left(  C_{2}\left(  \tau-\tau
_{0}\right)  \right)  }{\sqrt{\exp\left(  4\sqrt{6\left\vert E\right\vert
}\left(  \tau-\tau_{0}\right)  \right)  -4C_{2}^{2}}}%
\end{align*}
where $C_{1,2}=\varepsilon_{C}\left(  \frac{6\left\vert E\right\vert }%
{V_{1,2}}\right)  ^{\frac{1}{2}}$ , $\varepsilon_{C}=\pm1~$and from the
relation $dt=\sqrt{3}\left(  u^{2}-v^{2}\right)  d\tau$ we have%
\[
t-t_{0}=\sqrt{3}\left[  \ln\left(  \frac{2C_{1}e^{2C_{1}\tau_{0}}%
-e^{2C_{1}\tau}}{2C_{1}e^{2C_{1}\tau_{0}}+e^{2C_{1}\tau}}\right)  +\ln\left(
\frac{2C_{2}e^{2C_{2}\tau_{0}}+e^{2C_{2}\tau}}{2C_{2}e^{2C_{2}\tau_{0}%
}-e^{2C_{2}\tau}}\right)  \right]  .
\]

\subsubsection{Phantom, $\varepsilon=-1$}

In the case of phantom scalar field, i.e. $\varepsilon=-1$,$~\ $in parabolic
coordinates
\begin{equation}
a^{3}=\frac{3}{32}\left(  w^{2}+z^{2}\right)  ^{2}~\ ,~\phi=\sqrt{\frac{8}{3}%
}\arctan\left(  \frac{2wz}{w^{2}-z^{2}}\right)
\end{equation}
the Lagrangian (\ref{CS.05}) of the field equations becomes
\[
L\left(  w,z,\dot{w},\dot{z}\right)  =-\frac{\left(  w^{2}+z^{2}\right)  }%
{2}\left(  \dot{w}^{2}+\dot{z}^{2}\right)  +\frac{V_{1}w^{6}+V_{2}z^{6}}%
{u^{2}+v^{2}}.
\]
The field equations are the Hamiltonian%
\begin{equation}
-\frac{1}{\left(  w^{2}+z^{2}\right)  }\left[  \frac{1}{2}\left(  p_{w}%
^{2}+p_{z}^{2}\right)  -\frac{\bar{V}_{1}}{6}w^{6}-\frac{\bar{V}_{2}}{6}%
z^{6}\right]  =E \label{pha.02}%
\end{equation}
and Hamilton's equations of (\ref{pha.02}), where $p_{w},p_{z}$ are the
momenta and $\bar{V}_{1}=\frac{9}{4}\left(  c_{p1}+c_{p2}\right)  ,$ $\bar
{V}_{2}=\frac{9}{4}\left(  c_{p2}-c_{p1}\right)  .$

Working as previously we find that the solution of the Hamilton Jacobi
equation is
\[
S\left(  t,w,z\right)  =-\frac{\sqrt{3}}{3}\int\sqrt{\bar{V}_{1}%
w^{6}+6\left\vert E\right\vert w^{2}+\Phi_{0}}dw\pm\frac{\sqrt{3}}{3}\int
\sqrt{\bar{V}_{2}z^{6}+6\left\vert E\right\vert z^{2}-\Phi_{0}^{\prime}}dz-t.
\]

Therefore the reduced dynamical system is
\begin{align}
\left(  w^{2}+z^{2}\right)  \dot{w}  &  =\frac{\sqrt{3}}{3}\sqrt{\bar{V}%
_{1}w^{6}+6\left\vert E\right\vert w^{2}+\Phi_{0}^{\prime}}\label{pha.03}\\
\left(  w^{2}+z^{2}\right)  \dot{z}  &  =\pm\frac{\sqrt{3}}{3}\sqrt{\bar
{V}_{2}z^{6}+6\left\vert E\right\vert z^{2}-\Phi_{0}^{\prime}} \label{pha.04}%
\end{align}
The dynamical system (\ref{pha.03}), (\ref{pha.04}) is a two dimensional
nonlinear system. In order to simplify it we may apply the conformal
transformation $dt=\sqrt{3}\left(  w^{2}+z^{2}\right)  d\tau,~$i.e. $dt\cong
a^{\frac{3}{2}}d\tau,$ as we did in section \ref{quintsf}. Furthermore, from
the singularity condition $a\left(  t\rightarrow0^{+}\right)  \rightarrow
0^{+}$, we have $\left(  w,z\right)  _{t\rightarrow0}=0;$ hence,$~$ in order
to avoid complex solutions of the system (\ref{pha.03}), (\ref{pha.04}) we set
$\Phi_{0}^{\prime}=0$.

\section{Observational constraints}

\label{Cconstrain}

In this section, we test the viability of the cosmological model in the late
times (well inside the matter era) resulting for the scalar field potentials
we have determined, by performing a joint likelihood analysis using the SNIa,
BAO and the $H\left(  z\right)  $ data. The likelihood function is%
\begin{equation}
\mathcal{L}\left(  \mathbf{p}\right)  \mathcal{=L}_{SNIa}\mathcal{\times
L}_{BAO}\mathcal{\times L}_{H\left(  z\right)  }%
\end{equation}
where $\mathbf{p}$ is the statistical vector that contains the free parameters
and $\mathcal{L}_{A}\varpropto e^{-\chi_{A}^{2}/2}~$; that is, $\chi^{2}%
=\chi_{SNIa}^{2}+\chi_{BAO}^{2}+\chi_{H\left(  z\right)  }^{2}$.

For the Type Ia supernova data we use the Union 2.1 set which provides us with
580 SNIa distance modulus at observed redshift \cite{Suzuki}. The chi-square
is given by the expression\footnote{We have applied the diagonal covariant
matrix.}%
\begin{equation}
\chi_{SNIa}^{2}=\sum\limits_{i=1}^{N_{SNIa}}\left(  \frac{\mu_{obs}\left(
z_{i}\right)  -\mu_{th}\left(  z_{i};\mathbf{p}\right)  }{\sigma_{i}}\right)
^{2}%
\end{equation}
where $N_{SNIa}=580$, $z_{i}$ is the observed redshift $z_{i}\in$ $\left[
0.015,1.414\right]  $, $\mu$ is the distance modulus $\mu=m-M=5\log D_{L}+25$
and $D_{L}$ is the luminosity distance.

The chi-square for the Hubble parameter constraint data is%
\begin{equation}
\chi_{H\left(  z\right)  }^{2}=\sum\limits_{i=1}^{N_{H(z)}}\left(
\frac{H_{obs}\left(  z_{i}\right)  -H_{th}\left(  z_{i};\mathbf{p}\right)
}{\sigma_{i}}\right)  ^{2}%
\end{equation}
where $N_{H(z)}=21$, $H_{th}\left(  z_{i};\mathbf{p}\right)  $ is the
theoretical Hubble parameter and $H_{obs}$ are the 21 observed Hubble
parameters at the observed redshift $z_{i}$~\cite{simon,stern,Gaz,mar} (see
Table 1 of \cite{Farooq}).

Furthermore we use the 6dF, the SDSS and WiggleZ BAO data
\cite{Percival,BlakeC} for which the corresponding chi-square is%
\begin{equation}
\chi_{BAO}^{2}=\sum\limits_{i=1}^{N_{BAO}}\left(  \sum\limits_{j=1}^{N_{BAO}%
}\left[  d_{obs}\left(  z_{i}\right)  -d_{th}\left(  z_{i};\mathbf{p}\right)
\right]  C_{ij}^{-1}\left[  d_{obs}\left(  z_{j}\right)  -d_{th}\left(
z_{j};\mathbf{p}\right)  \right]  \right)
\end{equation}
where $N_{BAO}=6$, $C_{ij}^{-1}$ is the inverse of the covariant matrix in
terms of $d_{z}~\ $\cite{BasilNess}, and the parameter $d_{z}$ follows from
the relation $d_{z}=\frac{l_{BAO}}{D_{V}\left(  z\right)  }$; $~l_{BAO}\left(
z_{drag}\right)  $ is the BAO scale at the drag redshift and $D_{V}\left(
z\right)  $ is the volume distance \cite{BlakeC}.

Without losing the generality in the case of the quintessence UDM model we set
$\theta_{1}\rightarrow0$.
%scalar field with potential
%(\ref{Pot.1}) we make the ansatz $\left(  x_{0},\bar{x}_{0}\right)  =\left(
%y_{0},\bar{y}_{0}\right)  $,~ $c_{\left(  Q,p\right)  2}=$ $c_{\left(
%Q,p\right)  1}.~$Furthermore from fig. \ref{numer1} we observe that in order
%to have a behavor of $\Lambda$-cosmology the $\theta_{1}$ should be small and
%very close to zero; we select the values $\theta_{1}=0.01~$and$~\varepsilon
%_{\theta}=-1$.
Therefore, the UDM statistical vector $\mathbf{p~}$is of dimension two
$\left(  \dim\mathbf{p=2}\right)  ;$ that is, $\mathbf{p=}\left(  \omega
_{m0},c_{\left(  Q,p\right)  1}\right)  ,$ recall that $\omega_{m0}%
=\Omega_{m0}H_{0}^{2}$, and $\rho_{m0}=3\omega_{m0}=\left\vert E\right\vert .$
In order to constraint the new hyperbolic scalar field with potential
(\ref{Pot.2}) with the data, we have to define six free parameters, that is,
$c_{\left(  Q,p\right)  1},c_{\left(  Q,p\right)  2},\omega_{m0}$,~$\Phi
_{0}/\Phi_{0}^{^{\prime}}$ and the initial conditions$~\left(  u,v\right)
/\left(  w,z\right)  $. In order to reduce the number of the free parameters
we make the ansatz $c_{\left(  Q,p\right)  2}=0.8c_{\left(  Q,p\right)  1}.$
Furthermore from the singularity condition $a\left(  t\rightarrow0^{+}\right)
\rightarrow0^{+}$ we select the initial conditions $\left(  u,v\right)
_{t\rightarrow0^{+}}=\left(  0.1,0.0998\right)  ,~\left(  w,z\right)
_{_{t\rightarrow0^{+}}}=\left(  10^{-3},10^{-4}\right)  ~$where in
(\ref{St.2}) we considered the sign minus $"-"~$and in (\ref{pha.04}) the sign
plus~$"+"$. Finally we assume the integration constants $\Phi_{0}/\Phi
_{0}^{^{\prime}}~$to vanish .
%Similarly for the scalar field with potential
%(\ref{Pot.1}) we make the ansatz $\left(  x_{0},\bar{x}_{0}\right)  =\left(
%y_{0},\bar{y}_{0}\right)  $,~ $c_{\left(  Q,p\right)  2}=$ $c_{\left(
%Q,p\right)  1}.~$Furthermore from fig. \ref{numer1} we observe that in order
%to have a behavor of $\Lambda$-cosmology the $\theta_{1}$ should be small and
%very close to zero; we select the values $\theta_{1}=0.01~$and$~\varepsilon
%_{\theta}=-1$. Therefore, the constrain parameter\ $\mathbf{p~}$for the two
%scalar fields has dimension two $\left(  \dim\mathbf{p=2}\right)  ;$ that is,
%$\mathbf{p=}\left(  \omega_{m0},c_{\left(  Q,p\right)  1}\right)  ,$ recall
%that $\omega_{m0}=\Omega_{m0}H_{0}^{2}$, and $\rho_{m0}=3\omega_{m0}%
%=\left\vert E\right\vert .$

Lastly, since $N/n_{fit}>40$ we will use, the relevant to our case,
\emph{corrected} Akaike information criterion \cite{Akaike1974}, defined, for
the case of Gaussian errors, as:
\begin{equation}
\mathrm{AIC}=\chi_{min}^{2}+2n_{fit}%
\end{equation}
where $N=N_{SNIa}+N_{H(z)}+N_{BAO}=607$ and $n_{fit}$ is the number of free
parameters. A smaller value of \textrm{AIC} indicates a better model-data fit
(for the scalar field models $n_{fit}=2$ and for the $\Lambda$CDM we have
$n_{fit}=2$). However, small differences in AIC are not necessarily
significant and therefore, in order to assess, the effectiveness of the
different models in reproducing the data, one has to investigate the model
pair difference $\Delta$AIC$=\mathrm{AIC}_{y}-\mathrm{AIC}_{x}$. The higher
the value of $|\Delta\mathrm{AIC}|$, the higher the evidence against the model
with higher value of $\mathrm{AIC}$, with a difference $|\Delta$%
AIC$|\raise-3.truept\hbox{\rlap{\hbox{$\sim$}}\raise4.truept\hbox{$>$}\ }2$
indicating a positive such evidence and $|\Delta$%
AIC$|\raise-3.truept\hbox{\rlap{\hbox{$\sim$}}\raise4.truept\hbox{$>$}\ }6$
indicating a strong such evidence, while a value
$\raise-3.truept\hbox{\rlap{\hbox{$\sim$}}\raise4.truept\hbox{$<$}\ }2$
indicates consistency among the two comparison models.

The scalar field with potential (\ref{Pot.1}) has been compared with the
cosmological data in \cite{BasilLukes} for the quintessence field and in
\cite{capPhantom} for the phantom field. In contrast to \cite{BasilLukes} in
our solution we include the dark matter component in the field equations.
Furthermore in \cite{capPhantom} the authors examine the case where $\bar
{c}_{p1}=0~$and$~\bar{c}_{p2}\neq0$.%

%TCIMACRO{\TeXButton{B}{\begin{table}[tbp] \centering}}%
%BeginExpansion
\begin{table}[tbp] \centering
%EndExpansion
\caption{Best fit values and cosmological parameters for the SNIa+BAO and SNIa+BAO+H(z) tests
for the $\Lambda$-cosmology and for the Scalar fields dark energy models which admit dynamical symmetries. The final three column present the
goodness-of-fit statistics $\chi^{2}_{min}$ parameter}%
\begin{tabular}
[c]{cccccc}\hline\hline
$\Lambda$\textbf{CDM} & $\Omega_{m0}$ & $H_{0}$ & $w_{\Lambda}~$(fixed) &
$\omega_{m0}\times10^{-4}$ & $\chi_{min}^{2}$\\\hline
SNIa+BAO & $0.29_{-0.032}^{+0.016}$ & $68.4_{-1.40}^{+0.70}$ & $-1.000$ &
$0.14$ & $560.32$\\
SNIa+BAO+H(z) & $0.29_{-0.020}^{+0.016}$ & $68.4_{-0.10}^{+2.70}$ & $-1.000$ &
$0.14$ & $574.77$\\
&  &  &  &  & \\
\textbf{Scalar Field (\ref{Pot.1})} & $\Omega_{m0}$ & $H_{0}$ & $w_{\phi0}$ &
$\left(  \omega_{m0},c_{\left(  Q,p\right)  1}\right)  \times10^{-4}$ &
$\chi_{min}^{2}$\\\hline
\textit{Quintessence, }$\varepsilon=1$ &  &  &  &  & \\
SNIa+BAO & $0.25$ & $68.1$ & $-0.965$ & $\left(  0.12_{-0.019}^{+0.031}%
,0.39_{-0.052}^{+0.020}\right)  $ & $563.48$\\
SNIa+BAO+H(z) & $0.28$ & $69.8$ & $-0.968$ & $\left(  0.13_{-0.009}%
^{+0.015},0.39_{-0.028}^{+0.016}\right)  $ & $577.82$\\
\textit{Phantom, }$\varepsilon=-1$ &  &  &  &  & \\
SNIa+BAO & $0.29$ & $67.9$ & $-1.017$ & $\left(  0.13_{-0.021}^{+0.020}%
,0.58_{-0.048}^{+0.088}\right)  $ & $562.93$\\
SNIa+BAO+H(z) & $0.30$ & $69.6$ & $-1.016$ & $\left(  0.15_{-0.013}%
^{+0.008},0.60_{-0.012}^{+0.060}\right)  $ & $576.59$\\
&  &  &  &  & \\
\textbf{Scalar Field (\ref{Pot.2})} & $\Omega_{m0}$ & $H_{0}$ & $w_{\phi0}$ &
$\left(  \omega_{m0},c_{\left(  Q,p\right)  1}\right)  \times10^{-4}$ &
$\chi_{min}^{2}$\\\hline
\textit{Quintessence, }$\varepsilon=1$ &  &  &  &  & \\
SNIa+BAO & $0.27$ & $67.9$ & $-1.000$ & $\left(  0.12_{-0.014}^{+0.026}%
,0.14_{-0.010}^{+0.015}\right)  $ & $562.84$\\
SNIa+BAO+H(z) & $0.28$ & $69.7$ & $-1.000$ & $\left(  0.14_{-0.005}%
^{+0.015},0.15_{-0.010}^{+0.009}\right)  $ & $576.76$\\
\textit{Phantom, }$\varepsilon=-1$ &  &  &  &  & \\
SNIa+BAO & $0.27$ & $68.3$ & $-1.044$ & $\left(  0.13_{-0.017}^{+0.023}%
,0.15_{-0.012}^{+0.014}\right)  $ & $562.75$\\
SNIa+BAO+H(z) & $0.29$ & $69.6$ & $-1.048$ & $\left(  0.14_{-0.002}%
^{+0.009},0.15_{-0.013}^{+0.006}\right)  $ & $576.65$\\\hline\hline
\end{tabular}
\label{table2}%
%TCIMACRO{\TeXButton{E}{\end{table}}}%
%BeginExpansion
\end{table}%
%EndExpansion

In table \ref{table2} we give a numerical summary of the current statistical
analysis ~and the scalar field models with potentials (\ref{Pot.1}),
(\ref{Pot.2}). For the $\Lambda$-cosmology we find the minimum total
chi-square $\chi_{min}^{2}=574.77$ ($\mathrm{dof}=606$) with best fit values
$\left(  \Omega_{m0},H_{0}\right)  _{\Lambda}=\left(  0.29,68.4\right)  $. For
the scalar field model (\ref{Pot.1}) we find for the quintessence field
$\min_{Q1}\chi_{min}^{2}=577.82$ ($\mathrm{dof}=605$) and for the best fit
values of the parameters $\omega_{m0},c_{Q1}$ we have the cosmological
parameters $\left(  \Omega_{m0},H_{0},w_{\phi0}\right)  _{Q1}=\left(
0.28,69.8,-0.968\right)  $ whereas for the phantom field we have $\chi
_{min}^{2}=576.59$ and $\left(  \Omega_{m0},H_{0},w_{\phi0}\right)
_{P1}=\left(  0.30,69.6,-1.016\right)  .$

Similarly for the potential (\ref{Pot.2}) we find for the quintessence field
$\chi_{min}^{2}=576.76$,~ $\left(  \Omega_{m0},H_{0},w_{\phi0}\right)
_{Q2}=\left(  0.28,69.7,-1.000\right)  $ and for the phantom field $\chi
_{min}^{2}=576.65$,~ $\left(  \Omega_{m0},H_{0},w_{\phi0}\right)
_{P2}=\left(  0.29,69.6,-1.048\right)  .$

As it is expected the value of AIC$_{\Lambda}$($\simeq578.77$) is smaller than
the corresponding values of the scalar field models AIC$_{scalar}%
$($580.59-581.82$) which indicates that the $\Lambda$CDM model appears to fit
better than the scalar fields models the expansion data. However, the
differential value\footnote{For the quintessence UDM scalar field we find
$\left\vert \Delta AIC\right\vert \simeq3.1$.} $|\Delta\mathrm{AIC}%
|$=$|\mathrm{AIC}_{\Lambda}-\mathrm{AIC}_{scalar}|$ is actually $\leq2$ which
indicates that the cosmological data are perfectly consistent with the current
scalar field models in a way comparable to the concordance model.
%values (i.e.,
%$\sim1.8-2.1$) indicate that the expansion data are marginally consistent with
%the current scalar field models.

\begin{figure}[ptb]
\includegraphics[height=8cm]{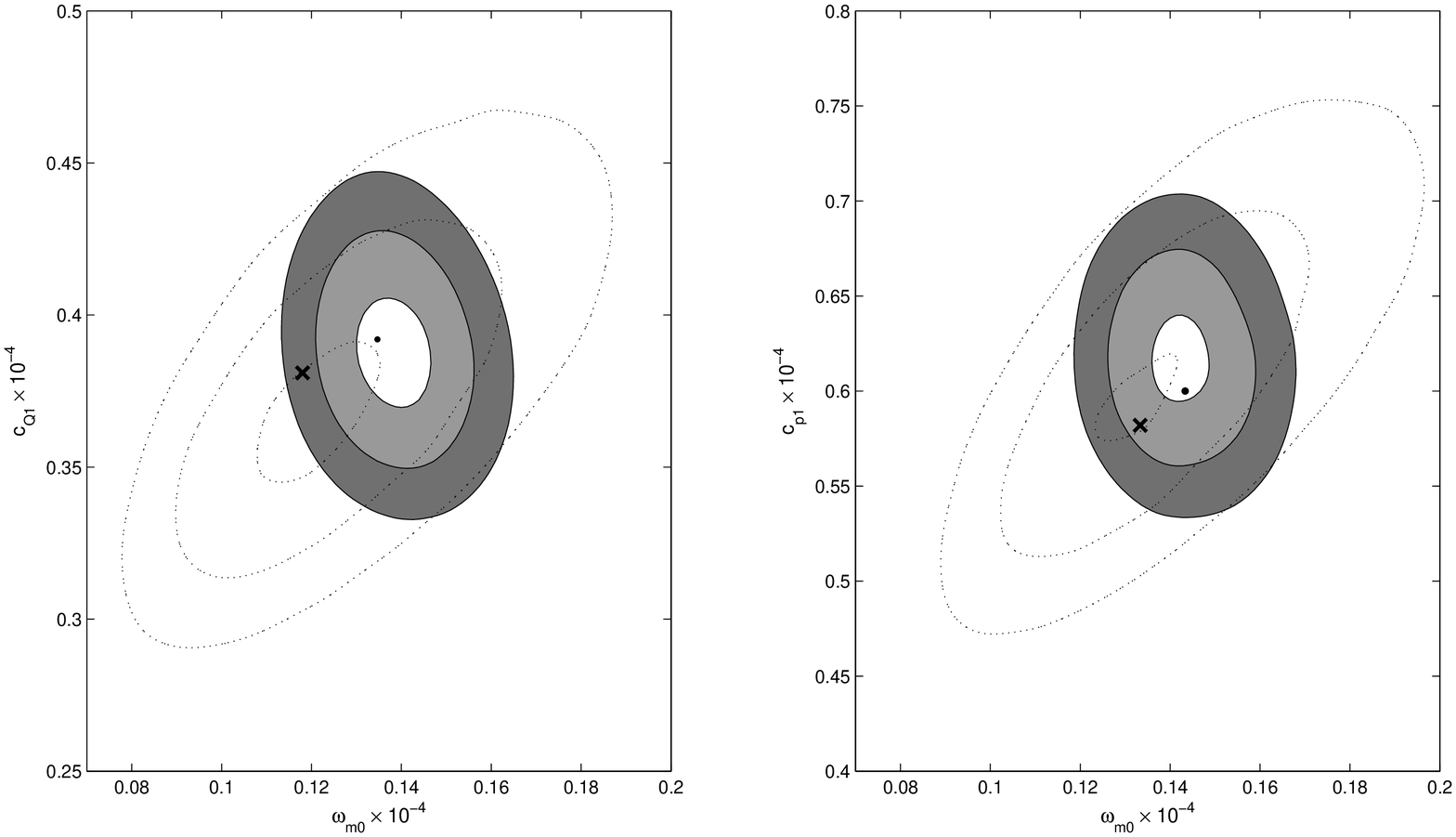}
%\mbox{\epsfxsize=14.2cm \epsffile{likelihoodModel1.eps}}
\caption{Likelihood contours of $1\sigma~\left(  \Delta\chi^{2}=2.3\right)
,2\sigma~\left(  \Delta\chi^{2}=6.18\right)  $ and $3\sigma~\left(  \Delta
\chi^{2}=11.83\right)  $ in the plane $\left(  \omega_{m0},c_{\left(
Q,p\right)  1}\right)  $ for the scalar field with potential (\ref{Pot.1}).
Left figure is for the Quintessence field\ whereas right figure is for the
Phantom field.~The filled areas are for the SnIa+BAO+H(z) test, the best fit
values are marked with a dot; the dot lines are for the SnIa+BAO test and the
best fit values are marked with $\times$. \ }%
\label{likcmod1}%
\end{figure}

\begin{figure}[ptb]
\includegraphics[height=8cm]{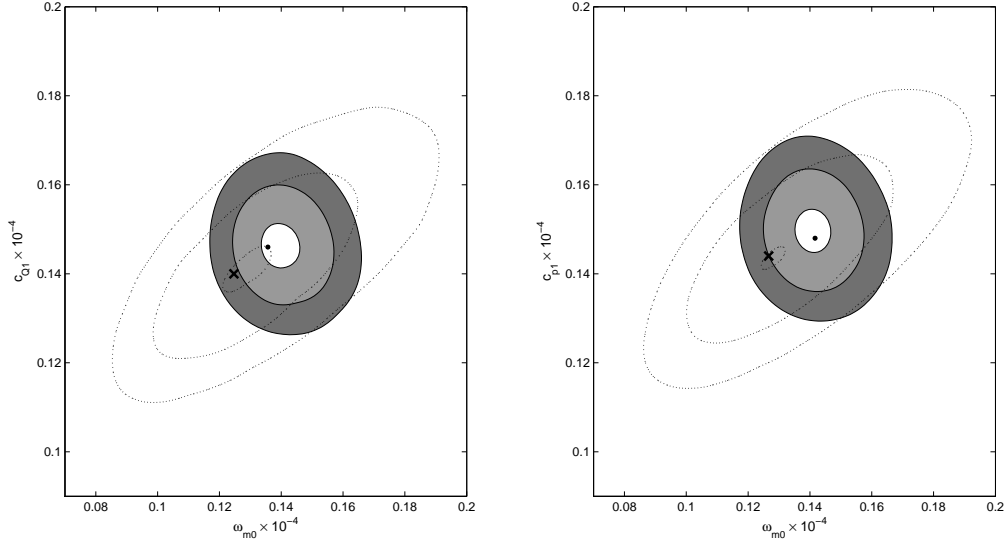}
%\mbox{\epsfxsize=14.2cm \epsffile{likelihoodModel1.eps}}
\caption{Likelihood contours of $1\sigma~\left(  \Delta\chi^{2}=2.3\right)
,2\sigma~\left(  \Delta\chi^{2}=6.18\right)  $ and $3\sigma~\left(  \Delta
\chi^{2}=11.83\right)  $ in the plane $\left(  \omega_{m0},c_{\left(
Q,p\right)  1}\right)  $ for the scalar field with potential (\ref{Pot.2}).
Left figure is for the Quintessence field\ whereas right figure is for the
Phantom field.~The filled areas are for the SnIa+BAO+H(z) test, the best fit
values are marked with a dot; the dot lines are for the SnIa+BAO test and the
best fit values are marked with $\times$. \ }%
\label{likcmod2}%
\end{figure}

In order to give the reader the opportunity to appreciate our observational
constraints, in fig. \ref{likcmod1} and \ref{likcmod2} we provide the
likelihood contours for the best fit parameters $\left(  \omega_{m0}%
,c_{\left(  Q,p\right)  1}\right)  $ of the scalar fields with potentials
(\ref{Pot.1}) and (\ref{Pot.2}).

%{\LARGE Moreover, the }$\Lambda-${\LARGE cosmology has the "frequency"
%}$_{\Lambda}\omega_{1}=\frac{3}{2}\sqrt{1-\Omega_{m}}\frac{H_{0}}{100}\frac
%{1}{9.78}=0.09~\left(  Gyrs\right)  ^{-1}${\LARGE and for the UDM potential
%(\ref{Pot.1})}$~${\LARGE has }$_{Q}\omega_{1}=\frac{\sqrt{\frac{3}{2}c_{Q1}}%
%}{9.78}=0.078~\left(  Gyrs\right)  ^{-1}${\LARGE and }$_{P}\omega_{1}%
%=\frac{\sqrt{\frac{3}{2}c_{P1}}}{9.78}=0.097~\left(  Gyrs\right)  ^{-1}.$

\section{Conclusion}

\label{Conclusion}

In this work we have applied the dynamical Noether symmetry approach as a
geometric rule to select the potential of the scalar field in the scalar field
cosmology. We have found two potentials with this property and have used the
resulting Noether integrals to integrate the corresponding field equations.
The first potential is the well known UDM model whereas the second potential
is new. In both cases we have found the analytic solution of the field
equations both in the quintessence and in the phantom case. The solution for
the new potential is expressed in terms of elliptic functions and contains a
number of free parameters. \ In order to find an explicit analytic solution we
consider certain simplifications which are compatible with the physical
assumptions. Furthermore we test the solutions we have found against the
observed cosmological constraints that is, the SNIa, BAO and the $H\left(
z\right)  $ data. We find that the cosmological parameters for the scalar
field models which admit dynamical symmetries are similar with those of the
$\Lambda$-cosmology

Besides the actual value of the new solution the approach shows that the use
of careful geometric requirements/ assumptions can help in two directions,
namely a.) to produce new results which is impossible to be found by ordinary
physical reasoning and b) to lead to models with a large number of free
parameters which provide adequate freedom of adjustment; in the sense that the
values of the constants are fixed in accordance with the observed data,
therefore leading to viable/ sound cosmological models.

\begin{acknowledgments}
AP acknowledges financial support of INFN (initiative specifiche QGSKY, QNP,
and TEONGRAV). SB acknowledges support by the Research Center for Astronomy of
the Academy of Athens in the context of the program \textit{\textquotedblright
Tracing the Cosmic Acceleration\textquotedblright}.
\end{acknowledgments}

\appendix

\section{Classification of scalar field potentials which admit dynamical
symmetries}

\label{appendix1}

In this appendix we give the complete classification of the potentials for
which the Lagrangian (\ref{CS.08}) admits contact Noether symmetries. We have
the following results.

\begin{itemize}
\item If the scalar field potential is
\begin{equation}
V\left(  \theta\right)  =c_{1}\left(  1-3\sinh^{2}\theta\right)  +c_{2}\left(
3\sinh\theta-\sinh^{3}\theta\right)
\end{equation}
Lagrangian (\ref{CS.08}) admits the additional dynamical symmetry%
\begin{equation}
X=-r^{2}\cosh\theta~\dot{\theta}\partial_{r}+\left(  \cosh\theta~\dot
{r}+2r\sinh\theta~\dot{\theta}\right)  \partial_{\theta}%
\end{equation}
with corresponding Noether Integral
\begin{equation}
\bar{I}_{2}=\left(  \cosh\theta~\dot{r}+r\sinh\theta~\dot{\theta}\right)
r^{2}\dot{\theta}-r^{3}\cosh^{2}\theta\left(  2c_{1}\sinh\theta-c_{2}\left(
1-\sinh^{2}\theta\right)  \right)  .
\end{equation}
This potential is equivalent to potential (\ref{Pot.2}) under the
transformation $\theta=\bar{\theta}+i\frac{\pi}{2}.$

\item If the scalar field potential is
\begin{equation}
V\left(  \theta\right)  =c_{1}+c_{2}e^{2\theta} \label{Pot3a}%
\end{equation}
Lagrangian (\ref{CS.08}) admits the dynamical symmetry%
\begin{equation}
X=-e^{2\theta}\left(  \dot{r}+r\dot{\theta}\right)  \partial_{r}%
+\frac{e^{2\theta}}{r}\left(  \dot{r}+r\dot{\theta}\right)  \partial_{\theta}%
\end{equation}
with corresponding Noether Integral
\begin{equation}
I_{3}=e^{2\theta}\left(  \left(  \dot{r}+r\dot{\theta}\right)  ^{2}%
-2r^{2}c_{1}\right)
\end{equation}
When $c_{1}=0$ the dynamical system admits the additional dynamical symmetry%
\begin{equation}
X_{3}=-r^{2}e^{\theta}\dot{\theta}\partial_{r}+e^{\theta}\left(  \dot
{r}+2r\dot{\theta}\right)  \partial_{\theta} \label{Pot4l}%
\end{equation}
with corresponding Noether Integral%
\begin{equation}
\bar{I}_{3}=r^{2}e^{\theta}\left(  \dot{r}\dot{\theta}+r\dot{\theta}%
^{2}\right)  +\frac{2}{3}c_{2}r^{3}e^{3\theta}.
\end{equation}

\item In the case where the potential is \footnote{The same result holds for
the case $\bar{\theta}=-\theta$.}
\begin{equation}
V\left(  \theta\right)  =c_{1}e^{2\theta}+c_{2}e^{3\theta} \label{Pot.4}%
\end{equation}
Lagrangian (\ref{CS.08}) admits the dynamical symmetry (\ref{Pot4l}) with
corresponding Noether Integral%
\begin{equation}
I_{3}=r^{2}e^{\theta}\left(  \dot{r}\dot{\theta}+r\dot{\theta}^{2}\right)
+r^{3}e^{3\theta}\left(  \frac{2}{3}c_{1}+c_{2}e^{\theta}\right)  .
\label{Pot.4II}%
\end{equation}
Note that the potential (\ref{Pot.4}) can be seen in the context of the early
dark energy potential \cite{Bar00}.
\end{itemize}

In order to complete our analysis in appendix \ref{appendb} we apply the
integral (\ref{Pot.4II}) in order to reduce the order of the field equations
with potential (\ref{Pot.4}).

\section{Reduction of order for the early dark energy potential with $\left(
\lambda_{1},\lambda_{2}\right)  =\left(  2,3\right)  .$}

\label{appendb}

In this appendix we reduce the field equations of the model with potential
(\ref{Pot.4}). In this case Lagrangian (\ref{CS.05}) becomes%
\begin{equation}
L\left(  a,\phi,\dot{a},\dot{\phi}\right)  =-3a\dot{a}^{2}+\frac{1}{2}%
a^{3}\dot{\phi}^{2}-a^{3}\left(  c_{1}e^{\frac{\sqrt{6}}{2}\phi}+c_{2}%
e^{\frac{3\sqrt{6}}{4}\phi}\right)  \label{ans.22}%
\end{equation}

We consider the coordinate transformation%
\[
a^{3}=\frac{3}{8}\xi^{2}\eta~,~\phi=\sqrt{\frac{8}{3}}\ln\left(  \frac
{\sqrt{\eta}}{\xi}\right)
\]
by which the Lagrangian (\ref{ans.22}) becomes%
\[
L\left(  \xi,\eta,\dot{\xi},\dot{\eta}\right)  =-\xi\dot{\eta}\dot{\xi}%
-\bar{c}_{1}\eta^{2}-\bar{c}_{2}\frac{\eta^{\frac{5}{2}}}{\xi}%
\]
where $\bar{c}_{1,2}=\frac{3}{8}c_{1,2}$. The Hamiltonian in normal
coordinates is
\begin{equation}
E=-\frac{p_{\xi}p_{\eta}}{\xi}+\bar{c}_{1}\eta^{2}+\bar{c}_{2}\frac
{\eta^{\frac{5}{2}}}{\xi} \label{ans.23}%
\end{equation}

The field equations are the Hamiltonian constraint (\ref{ans.23}) and
Hamilton's equations%
\[
\xi\dot{\eta}=-p_{\xi}~~,~~\xi\dot{\xi}=-p_{\eta}%
\]%
\[
\dot{p}_{\xi}=\frac{c_{2}\eta^{\frac{5}{2}}-p_{\xi}p_{\eta}}{\xi^{2}}%
~~,~~\dot{p}_{\eta}=-\left(  2\bar{c}_{1}\eta+\frac{5}{2}\bar{c}_{2}\frac
{\eta^{\frac{3}{2}}}{\xi}\right)  .
\]

We note that (\ref{ans.23}) is in the form of equations (16) and (18) of
\cite{Daskal} (where $F\left(  \eta\right)  =1,~G\left(  \eta\right)
=0,~f\left(  \eta\right)  =\bar{c}_{1}\eta^{2}$ and $g\left(  \eta\right)
=\bar{c}_{2}\eta^{\frac{5}{2}}$). The solution of the Hamilton Jacobi
equation
\[
-\frac{1}{\xi}\left(  \frac{\partial S}{\partial\xi}\right)  \left(
\frac{\partial S}{\partial\eta}\right)  +\bar{c}_{1}\eta^{2}+\bar{c}_{2}%
\frac{\eta^{\frac{5}{2}}}{\xi}-E\frac{\partial S}{\partial t}=0
\]
is%
\[
S\left(  t,\xi,\eta\right)  =-\frac{\xi}{3}\sqrt{6\bar{c}_{1}\eta
^{3}+18\left\vert E\right\vert ~\eta+\Phi_{0}}-\int\frac{3\bar{c}_{2}%
\eta^{\frac{5}{2}}}{\sqrt{6\bar{c}_{1}\eta^{3}+18\left\vert E\right\vert
~\eta+\bar{\Phi}_{0}}}d\eta-t.
\]
where $\bar{\Phi}_{0}\propto I_{3}$.

Therefore the reduced Hamilton's equations are%
\begin{align}
\xi\dot{\eta}  &  =\frac{1}{3}\sqrt{6\bar{c}_{1}\eta^{3}+18\left\vert
E\right\vert \eta+\bar{\Phi}_{0}}\label{ans.24}\\
~~\xi\dot{\xi}  &  =\frac{3\left(  \left(  \bar{c}_{1}\eta^{2}+\left\vert
E\right\vert \right)  \xi+\bar{c}_{2}\eta^{\frac{5}{2}}\right)  }{\sqrt
{6\bar{c}_{1}\eta^{3}+18\left\vert E\right\vert ~\eta+\bar{\Phi}_{0}}.}
\label{ans.25}%
\end{align}

\end{document}